\documentclass[journal]{IEEEtran}

\usepackage{mathtools,amssymb,lipsum}
\usepackage{cuted}
\makeatletter
\newcommand{\thickhline}{%
    \noalign {\ifnum 0=`}\fi \hrule height 1pt
    \futurelet \reserved@a \@xhline
}
\makeatother

\usepackage{multirow}
\usepackage{dblfloatfix}
\usepackage{algorithm}
\usepackage{algpseudocode}
\usepackage{hyperref}
\usepackage{multirow}

\usepackage[caption=false, font=footnotesize]{subfig} %\captionsetup[sub]{font=small}

\algnewcommand\algorithmicswitch{\textbf{switch}}
\algnewcommand\algorithmiccase{\textbf{case}}
\algnewcommand\algorithmicassert{\texttt{assert}}
\algnewcommand\Assert[1]{\Stabhte \algorithmicassert(#1)}%
\algdef{SE}[SWITCH]{Switch}{EndSwitch}[1]{\algorithmicswitch\ #1\ \algorithmicdo}{\algorithmicend\ \algorithmicswitch}%
\algdef{SE}[CASE]{Case}{EndCase}[1]{\algorithmiccase\ #1}{\algorithmicend\ \algorithmiccase}%
\algtext*{EndSwitch}%
\algtext*{EndCase}%

\usepackage{multicol}
\usepackage{amsmath}
\usepackage{array}
\usepackage{makecell}
\usepackage{stackengine}
\newcommand\xrowht[2][0]{\addstackgap[.5\dimexpr#2\relax]{\vphantom{#1}}}
\newcolumntype{P}[1]{>{\centering\arraybackslash}p{#1}}

\begin{document}
%
% paper title
% Titles are generally capitalized except for words such as a, an, and, as,
% at, but, by, for, in, nor, of, on, or, the, to and up, which are usually
% not capitalized unless they are the first or last word of the title.
% Linebreaks \\ can be used within to get better formatting as desired.
% Do not put math or special symbols in the title.
% \title{Sparse and Semantic Representation of\\Physical Environments with 3D Scene Graphs\\for Intelligent Agents}
\title{I-Keyboard: Fully Imaginary Keyboard on Touch Devices Empowered by Deep Neural Decoder}
% Design of a Memory Architecture with a Feedback Mechanism/modulation for Service Provision by Robot and IoT Collaboration

% Service provision by robot and iot collaboration using a memory architecture with a feedback mechanism/modulation

%
% author names and IEEE memberships
% note positions of commas and nonbreaking spaces ( ~ ) LaTeX will not break
% a structure at a ~ so this keeps an author's name from being broken across
% two lines.
% use \thanks{} to gain access to the first footnote area
% a separate \thanks must be used for each paragraph as LaTeX2e's \thanks
% was not built to handle multiple paragraphs
%

\author{Ue-Hwan Kim, Sahng-Min Yoo and
        Jong-Hwan Kim,~\IEEEmembership{Fellow,~IEEE}% <-this % stops a space
%\thanks{Manuscript received April 14, 2018; accepted May 2, 2018. Date of publication May 2, 2018; date of current version May 15, 2018. This paper was recommended by Associate Editor A. F. S. Gomez. (Corresponding author: Jong-Hwan Kim.)}
\thanks{This work was supported by the National Research Foundation of Korea (NRF) grant funded by the Korea government (MSIT) (No. NRF-2017R1A2A1A17069837).
}
\thanks{The authors are with the School of Electrical Engineering, KAIST (Korea Advanced Institute of Science and Technology), Daejeon, 34141, Republic of Korea (e-mail: {\{uhkim, smyoo, johkim\}@rit.kaist.ac.kr}).}% <-this % stops a space
%\thanks{Jong-Hwan Kim is with the Faculty of the School of Electrical Engineering, Korea Advanced Institute of Science and Technology (KAIST), Daejeon, 34141, South Korea
%        {\tt\small johkim@rit.kaist.ac.kr}}% <-this % stops a space
%\thanks{Color versions of one or more of the figures in this paper are available online at http://ieeexplore.ieee.org.}
%\thanks{Digital Object Identifier 10.1109/TCYB.2017.2691666}
}

% note the % following the last \IEEEmembership and also \thanks - 
% these prevent an unwanted space from occurring between the last author name
% and the end of the author line. i.e., if you had this:
% 
% \author{....lastname \thanks{...} \thanks{...} }
%                     ^------------^------------^----Do not want these spaces!
%
% a space would be appended to the last name and could cause every name on that
% line to be shifted left slightly. This is one of those "LaTeX things". For
% instance, "\textbf{A} \textbf{B}" will typeset as "A B" not "AB". To get
% "AB" then you have to do: "\textbf{A}\textbf{B}"
% \thanks is no different in this regard, so shield the last } of each \thanks
% that ends a line with a % and do not let a space in before the next \thanks.
% Spaces after \IEEEmembership other than the last one are OK (and needed) as
% you are supposed to have spaces between the names. For what it is worth,
% this is a minor point as most people would not even notice if the said evil
% space somehow managed to creep in.

% The paper headers
\markboth{IEEE Transactions on Cybernetics}%
{Shell \MakeLowercase{\textit{et al.}}: Bare Demo of IEEEtran.cls for IEEE Journals}

% If you want to put a publisher's ID mark on the page you can do it like
% this:
%\IEEEpubid{0000--0000/00\$00.00~\copyright~2015 IEEE}
% Remember, if you use this you must call \IEEEpubidadjcol in the second
% column for its text to clear the IEEEpubid mark.

% use for special paper notices
%\IEEEspecialpapernotice{(Invited Paper)}

% make the title area
\maketitle

% As a general rule, do not put math, special symbols or citations
% in the abstract or keywords.
\begin{abstract}
Text-entry aims to provide an effective and efficient pathway for humans to deliver their messages to computers. With the advent of mobile computing, the recent focus of text-entry research has moved from physical keyboards to soft keyboards. Current soft keyboards, however, increase the typo rate due to lack of tactile feedback and degrade the usability of mobile devices due to their large portion on screens. To tackle these limitations, we propose a fully imaginary keyboard (I-Keyboard) with a deep neural decoder (DND). The invisibility of I-Keyboard maximizes the usability of mobile devices and DND empowered by a deep neural architecture allows users to start typing from any position on the touch screens at any angle. To the best of our knowledge, the eyes-free ten-finger typing scenario of I-Keyboard which does not necessitate both a calibration step and a pre-defined region for typing is first explored in this work. For the purpose of training DND, we collected the largest user data in the process of developing I-Keyboard. We verified the performance of the proposed I-Keyboard and DND by conducting a series of comprehensive simulations and experiments under various conditions. I-Keyboard showed 18.95\% and 4.06\% increases in typing speed (45.57 WPM) and accuracy (95.84\%), respectively over the baseline.
\end{abstract}

% Note that keywords are not normally used for peerreview papers.

% For peer review papers, you can put extra information on the cover
% page as needed:
% \ifCLASSOPTIONpeerreview
% \begin{center} \bfseries EDICS Category: 3-BBND \end{center}
% \fi
%
% For peerreview papers, this IEEEtran command inserts a page break and
% creates the second title. It will be ignored for other modes.
\IEEEpeerreviewmaketitle

\section{Introduction}
% The very first letter is a 2 line initial drop letter followed
% by the rest of the first word in caps.
% 
% form to use if the first word consists of a single letter:
% \IEEEPARstart{A}{demo} file is ....
% 
% form to use if you need the single drop letter followed by
% normal text (unknown if ever used by the IEEE):
% \IEEEPARstart{A}{}demo file is ....
% 
% Some journals put the first two words in caps:
% \IEEEPARstart{T}{his demo} file is ....
% 
% Here we have the typical use of a "T" for an initial drop letter
% and "HIS" in caps to complete the first word.
\IEEEPARstart{T}{ext}-entry takes a crucial role in human-computer interaction (HCI) applications \cite{kristensson2015next}. It is one of the most effective and efficient methods for humans to deliver messages to computers. In the early stage of HCI research, physical keyboard-based text-entry methods were prevailing. Since then, researchers have focused on designing keyboards with high usability. In the post-PC era, the birth of mobile \cite{abbas2018mobile} and ubiquitous computing have prompted the development of soft keyboards. Soft keyboards \cite{kumar2016user} set mobile devices free from equipping additional hardware, thus improve the mobility of mobile devices.

However, contemporary soft keyboards possess a few limitations. In fact, current soft keyboard techniques damage the usability of mobile devices in multiple ways other than the mobility. First, the lack of tactile feedback increases the rate of typos. The absence of tactile feedback causes hand drifts and tap variability among people when typing with soft keyboards \cite{li2013effects}. Users gradually misalign their hands and touch different points for the same input without the physical boundaries. In addition, users type in an eyes-free manner \cite{raman2008eyes} in a couple of mobile computing context where a display is separated from the typing interface. In such situations, the effect of hand drift becomes greater.

Second, soft keyboards hinder mobile devices from presenting enough content because they occupy a relatively large portion on displays. Mobile devices provide smaller displays than non-mobile devices in general and soft keyboards can fill up to 40\% of displays. A survey over 50 iPad users \cite{li20111line} indicates that users show dissatisfaction towards current soft keyboard dimensions. Users wish to minimize the size of virtual keyboard to view more information from the display. In short, we claim that current soft keyboards on mobile devices degrade the usability of mobile screens.

In this paper, we propose the imaginary keyboard (I-Keyboard) along with the deep learning (DL) based decoding algorithm to tackle the above-mentioned limitations of soft keyboards. First of all, the proposed I-Keyboard is invisible, which maximizes the utility of the screens on mobile devices. Users can view the content of an application in full screen and type freely at the same time. To further improve usability, I-Keyboard does not have pre-defined layout, shape or size of keys. Users can start typing on any position at any angle on touch screens without worrying about the keyboard position and shape.

Next, I-Keyboard comes with DL-based decoding algorithm which does not require a calibration step. The proposed deep neural decoder (DND) effectively handles both hand-drift and tap variability and dynamically translates the touch points into words. In order to design DND for I-Keyboard, we conduct a user study to analyze user behaviors and collect data to train DND. To the best of our knowledge, this data is the largest dataset for designing a soft keyboard decoder. For training DND, we formulate an auxiliary loss without which deep architectures does not converge. Moreover, the eyes-free ten-finger typing scenario which not only discards a calibration step, but also liberates users from typing at a pre-defined area has not been explored. The proposed I-Keyboard attempts to unfold such truly imaginary soft keyboard for the first time.

In summary, the main contributions of our work are as follows.
\begin{enumerate}
\item \textbf{Research Scenario}: We define an advanced typing scenario where both a pre-defined typing area and a calibration step are omitted. 
\item \textbf{Data Collection}: We collect user data in an unconstrained environment and comprehensively analyze user behaviors in such an environment.
\item \textbf{Deep Neural Decoder}: We design a DL architecture for a shape, position and angle independent decoding and formulate the auxiliary loss for training DND. % We propose a shape, position and angle independent decoding algorithm using deep learning techniques.
\item \textbf{Verification}: We verify the effectiveness of the proposed I-Keyboard through extensive simulations and experiments.
\item \textbf{Open Source}: We make the materials developed in this work public\footnote{\url{https://github.com/Uehwan/I-Keyboard}}: the source code for the data collection tool, the collected data, the data preprocessing tool and the I-Keyboard framework.
\end{enumerate}

The rest of this paper is structured as follows. In Section II, we review conventional research approaches and compare them with the proposed I-Keyboard. Section III describes the data collection process and analyzes user behaviors. Section IV delineates the I-Keyboard system architecture and DND. Sections V and VI verify the performance of I-Keyboard through comprehensive simulations and experiments. Discussion points follow in Section VII and concluding remarks in Section VIII. 

% needed in second column of first page if using \IEEEpubid
%\IEEEpubidadjcol

\section{Related Work}
In this section, we review previous research outcomes relevant to the proposed I-Keyboard. We discuss the main ideas and limitations of previous works and compare them with the proposed text-entry system.

\subsection{Intelligent Text-Entry}
Intelligent text-entry aims to provide quick and accurate typing interfaces to users. Although mechanical keyboards can also employ intelligent text-entry schemes, the schemes in general target virtual keyboards since virtual keyboards offer simpler ways to modify the keyboards. In this paper, we categorize intelligent text-entry schemes into three classes: gesture-based, optimized, and imaginary text-entries.

First of all, gesture-based text-entry allows drawing-like typing \cite{fuccella2014novice, abuhmed2015uoit}. Drawing-like typing removes the need for localizing each key position and users can start drawing from any place on the screen in an eyes-free manner. Though gesture-based text-entry offers concise eyes-free typing interfaces, it requires gesture recognition algorithms, which can hardly achieve a high accuracy \cite{poularakis2016low}. Gesture variability among users and similarities between gestures for each key cause ambiguity that increases the inherent difficulty of sequence classification \cite{devanne20153}. In addition, gesture-based text-entry takes longer time than other text-entry methods since each key involves a gesture rather than a touch or a key press. The proposed I-Keyboard targets a more tractable decoding problem and utilizes deep learning techniques to successfully deal with the variability among users.

For the second point, optimized text-entry supplies accessible and comfortable typing interfaces by optimizing the size, shape, and position of keys \cite{yang2016modifying, sarcar2018ability}. Current optimized text-entry methods require users to learn new typing interfaces \cite{ushida2014ippitsu} because knowledge transfer seldom occurs for novel typing interfaces. Furthermore, the optimization process frequently demands a calibration step. The calibration step complicates the usage of the optimized text-entries. I-Keyboard proposed in this paper does not involve learning and calibration processes since it operates with ten fingers and its decoding algorithm does not need any prior knowledge. 

Last but not least, imaginary keyboards, which are invisible to users, save invaluable screen resources and enables multi-tasking in the context of mobile computing \cite{gupta2016porous}. Imaginary keyboards reduce constraints during interaction and users can freely and comfortably deliver their messages. In addition, the imaginary keyboards coincide with the potent vision for user interfaces (UI) which has evolved from mechanical, graphical and gestural UI to imaginary UI, achieving tighter embodiment and more directness \cite{fishkin1999embodied}. Conventional works on imaginary UI, however, have only shown the feasibility not reaching the practical deployment level. Our I-Keyboard proposes a new concrete concept for imaginary UI and demonstrates the practical implementation of the concept deployed in a real-world environment.

\subsection{Ten Finger Typing}
Ten finger typing is one of the most natural and common text-entry methods \cite{lyons2004expert}. Users can achieve typing speed of 60 - 100 words per minute (WPM) by ten finger typing on physical keyboards \cite{shi2018toast}. Ten finger typing experience stored in muscle memory and tactile feedback from mechanical keys enable eyes-free typing \cite{clawson2005impacts}. However, transferring ten-finger typing knowledge from mechanical keyboards to soft keyboards does not successfully occur in general due to lack of tactile feedback, though a few works have shown the viability in special use cases \cite{rosenberg1999chording, almusaly2018evaluation}.

A number of works have attempted to understand the user behavior with ten finger typing with soft keyboards. The major findings are as follows:
\begin{enumerate}
    \item \textbf{Speed}: The typing speed with soft keyboards drops dramatically compared to the speed with physical keyboards \cite{mackenzie1999text}.
    \item \textbf{Typing pattern}: The distribution of touch points resembles the mechanical keyboard layout \cite{findlater2011typing}, though the distribution varies over time in shape and size \cite{shi2018toast}.
    \item \textbf{Hand drift}: Hand-drift occurs over time and becomes stronger for invisible keyboards \cite{li2013effects}.
    \item \textbf{Tap variability}: Various factors cause tap variability among users \cite{lu2017blindtype}. The factors include finger volume, hand posture and mobility.
\end{enumerate}

In summary, ten-finger eyes-free typing on virtual keyboards, which is most natural and easy to transfer knowledge directly from physical keyboards \cite{shi2018toast}, is feasible according to the previous research results, though a couple of obstacles need to be resolved. The proposed DND handles hand drift, tab variability and automatic calibration with a deep neural architecture to improve typing speed and to reduce error rate.

\subsection{Decoding Algorithms}
Classical statistical decoding algorithms translate user inputs (key strokes) into characters or words using probabilistic models. These statistical decoding algorithms have proved their effectiveness in a few controlled environments \cite{yi2017too}. The goal of statistical decoding is to find the sequence of characters that maximizes the joint probability of the given user input sequence. Mathematically, the user typing pattern for the decoding process is formulated as
\begin{equation}
    \hat{\mathbb{C}} = \mathop {\arg \max }\limits_{\mathbb{C}} {P(t_{1}, ..., t_{n}|c_{1}, ..., c_{n})},
    \label{eq:decoding_process}
\end{equation}
where $t_i$'s are the position of each key stroke, $\hat{\mathbb{C}} = (\hat{c}_{1}, ..., \hat{c}_{n})$, $c_{i}$'s are the characters and $n$ is the length of the sequence. Since the complexity of modeling the joint probability becomes untractable as the sequence length increases, the independence assumption is employed in most cases. By assuming the independence property, (\ref{eq:decoding_process}) becomes
\begin{equation}
    \hat{\mathbb{C}} \simeq \mathop {\arg \max }\limits_{\mathbb{C}} {\prod_{i=1}^{n} P(t_{i}|c_{i})}.
    \label{eq:decoding_process_simplified}
\end{equation}
The probability $P(t_{i}|c_{i})$ is approximated by a Gaussian distribution with Markov-Bayesian algorithm \cite{shi2018toast} or a bivariate Gaussian distribution \cite{zhu2018typing} in conventional approaches. In addition, the probability is separately modelled for left and right hands.

Conventional statistical decoding algorithms, however, cannot perfectly deal with the complex dynamics of user inputs. The independence assumption applied in these methods cannot count both long-term and short-term dependencies among the key strokes. The independence assumption confines the conventional approaches to regard only the current input. Furthermore, the previous research outcomes have proposed fixed models for statistical decoding, thus they cannot adaptively handle hand drift and tap variability that vary over time. Though a few have designed adaptive models, those models require either an additional calibration step or a controlled experiment environment \cite{shi2018toast}.

On the other hand, the proposed I-Keyboard comes with DND that overcomes the limitations of previous approaches. We train DND to translate the user input data containing hand-drifts and tap variability into a sequence of characters. The deep neural architecture for DND inherently models the long-term dependency without relying on the independence assumption and its implicit dynamic feature successfully eliminates the effect of hand drift and tap variability. Moreover, the semantic embedding integrated in the neural architecture functions as a language model that boosts the decoding performance. The reduced computational cost due to the joint architecture of decoding and language models is another advantage.

\section{User Study: Understanding User Behavior}
In order to understand the user behaviors on invisible and layout-free keyboards, we design a user study. Specifically, we aim to answer three questions: 1) ``can users type on touch screens without any visual and physical guidances from the screens?", 2) ``if they could, what are the characteristics of their typing behaviors in such an environment?" and 3) ``what type of feedback could improve the user experience during data collection?"

\subsection{Data Collection}
\subsubsection{Participants}
We recruited 43 participants from the campus (32 male, 11 female) aged from 22 to 32 (average = 25.84, std = 2.68). All participants regularly use both physical QWERTY keyboards and soft keyboards on touchable mobile devices. All of them could type in an eyes-free manner with physical keyboards.
\subsubsection{Apparatus}
\begin{figure}
	\centering
	\includegraphics[scale=0.5]{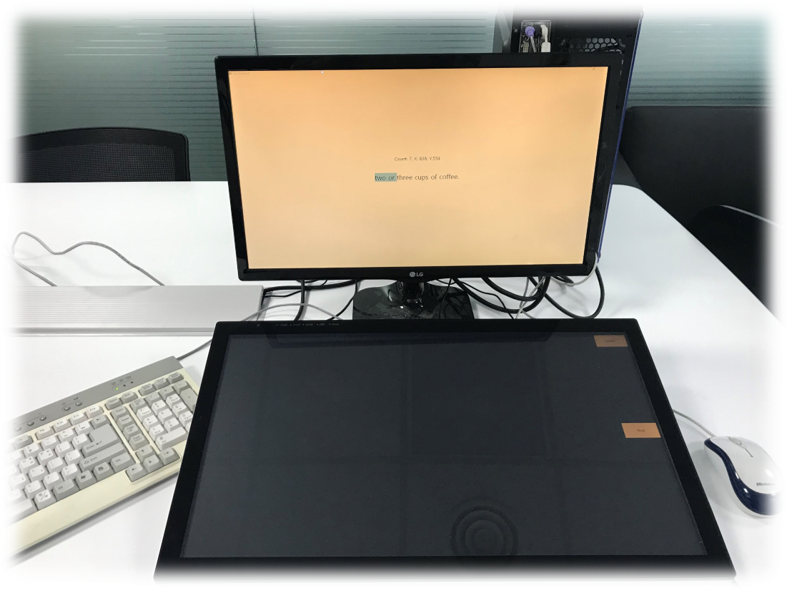}
	\caption{The data collection environment. We set up an ordinary monitor and a touch screen in a room.}
	\label{fig:data_collection_env}
\end{figure}
Fig. \ref{fig:data_collection_env} shows the overall data collection environment. We set up one touch screen (LG 23ET63, $1920 \times 1080$ pixels, $555\times 338$ mm$^{2}$, touchable) for text-entry and one ordinary screen (LG L1980QY, $1280 \times 1024$ pixels, $422\times 410$ mm$^{2}$, non-touchable) for instruction on a desk in a room for data collection. We placed the touch screen parallel to the desk so that users could treat the screen as a usual keyboard. We placed the ordinary screen for instruction perpendicular to the desk.
\subsubsection{Typing Interface}
\begin{figure}
    \centering
    \subfloat[Interface of the instruction screen.]{
        \includegraphics[width=0.22\textwidth]{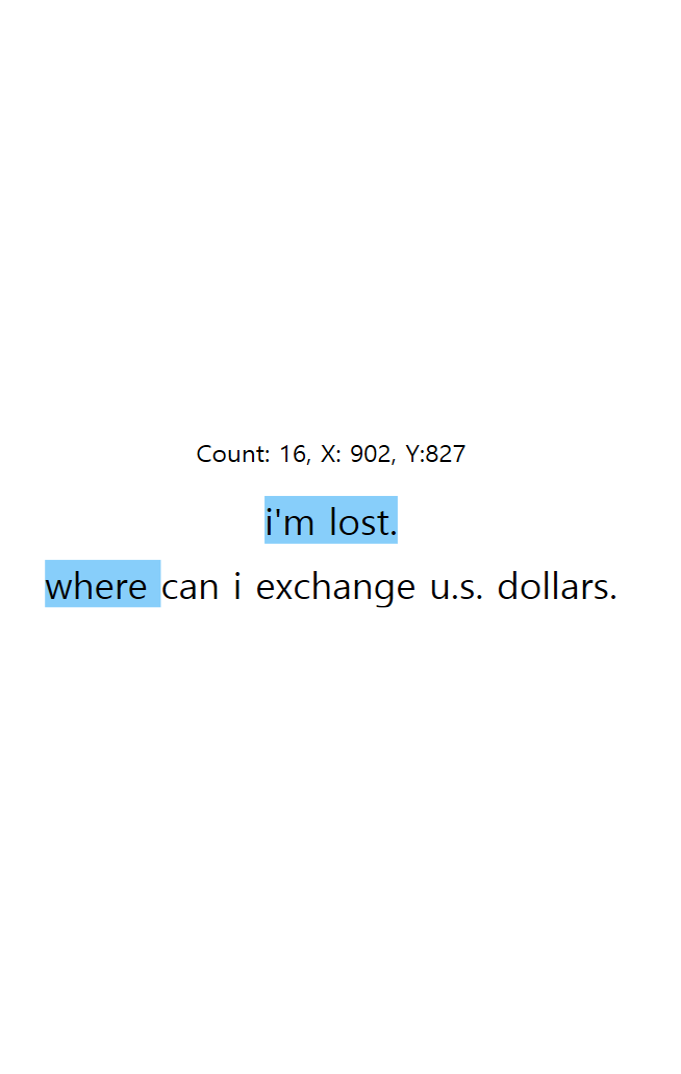}
        }
    \label{fig:instruction_screen}
    ~ %add desired spacing between images, e. g. ~, \quad, \qquad, \hfill etc. 
      %(or a blank line to force the subfigure onto a new line)
    \subfloat[Interface of the touch screen.]{
        \includegraphics[width=0.22\textwidth]{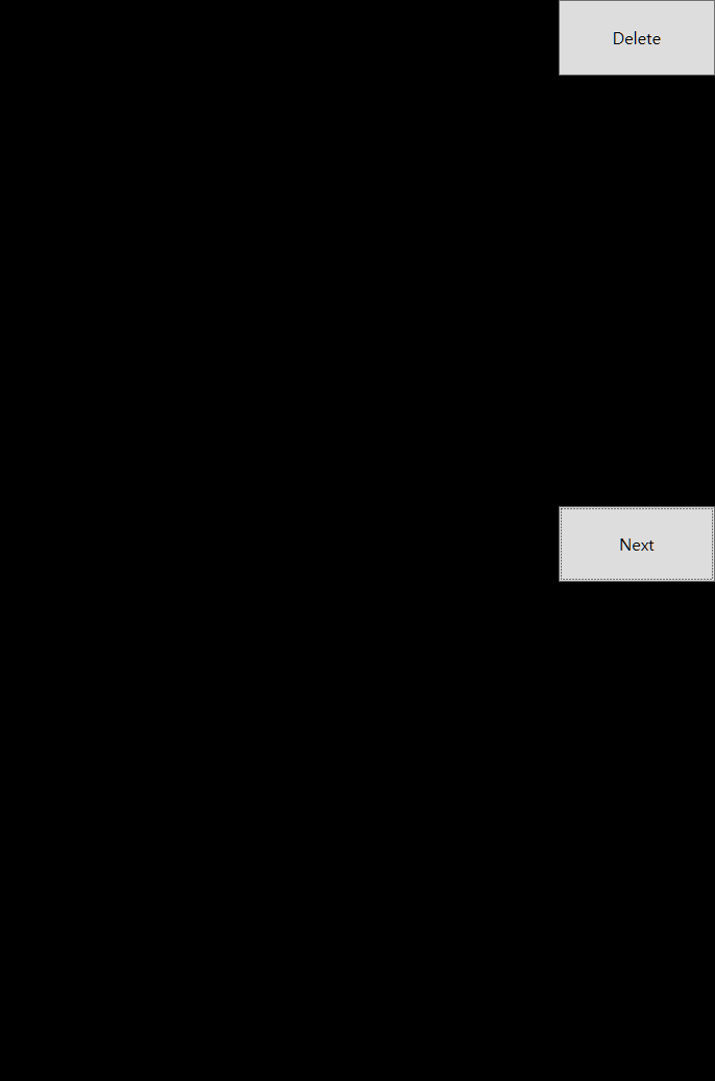}
        }
    \label{fig:touch_screen}
    ~ %add desired spacing between images, e. g. ~, \quad, \qquad, \hfill etc. 
    %(or a blank line to force the subfigure onto a new line)
    \caption{Typing interface. Only the center part of the instruction screen and the right half part of the touch screen are shown.}
    \label{fig:typing_interface}
\end{figure}
Fig. \ref{fig:typing_interface} illustrates the interface of the instruction and the typing screens. We implemented the interface using Windows Presentation Foundation (WPF). The instruction screen shows the task sentence (phrase) at the center. During typing, three types of feedback could be offered: none, asterisk \cite{shi2018toast, zhu2018typing} and highlight. For highlight feedback which is evaluated in this study for the first time, each character of the task sentence gets background-highlighted step by step as a legal touch is detected. The highlight feedback ensures one-to-one mapping between the touch points and the characters in the task sentence.

Next, users could start typing from anywhere on the typing screen at any angle. Neither a pre-determined area nor constraints were set for typing. The typing screen only contains the proceed and the delete buttons. Completing the current task sentence activates the proceed button and users can touch no more than the length of the task sentence. Users can delete the touch points collected for the current sentence in case users feel they have made typing mistakes.

\subsubsection{Procedure}
We first collected the demographics of participants and instructed the participants the data collection procedure. We asked users to assume there exists a keyboard on the touch screen and type as naturally as possible. For the first 15 sentences, participants familiarized themselves with the typing interface. After the warm-up, users transcribed 150 - 160 sentences randomly sampled from Twitter and 20 Newsgroup datasets. We preprocessed the task sentences so that the sentences only included 26 English letters in lower case, enter, space, period (.) and apostrophe ('). We randomly connected two sentences with the enter key. The data collection process for each participant took approximately 50 minutes. In total, the data collection included 7,245 phrases and 196,194 key stroke instances. Table \ref{tb:dataset_comparison} compares the dataset size collected in this study with those of other studies and verifies that the dataset is the largest among the recent studies.

In addition, we randomly selected nine of the participants and asked them to evaluate the three types of feedback. The selected participants typed ten phrases with each feedback, totalling 30 phrases. After the typing session, we collected three subjective ratings from the participants for each feedback type in 5-Likert scale: intuitiveness, convenience and overall satisfaction.

\begin{table}
\centering
\caption{Comparison of Dataset Size}
\def\arraystretch{1.3}
\begin{tabular}{c | c | c | c | c}
\hline
\thickhline
 & Participants & Age & Phrases & Points \\
\thickhline\xrowht{12pt}
BlindType \cite{lu2017blindtype} & $12$ & $20-32$ & \makecell{$9,664$ \\ (words)} & $40,509$ \\
\hline\xrowht{12pt}
TOAST \cite{shi2018toast} & $15$ & $22$ (mean) & $450$ & $12,669$ \\
\hline\xrowht{12pt}
\makecell{Invisible \\Keyboard \cite{zhu2018typing}} & $18$ & $18-50$ & $2,160$ & N.A.\\
\hline\xrowht{12pt}
Ours & $43$ & $25$ (mean) & $7,245$ & $196,194$ \\

\hline
\thickhline
\end{tabular}
\label{tb:dataset_comparison}
\end{table}

\subsection{User Behavior Analysis}
\subsubsection{User Mental Model}
\begin{figure}
	\centering
	\includegraphics[scale=0.2]{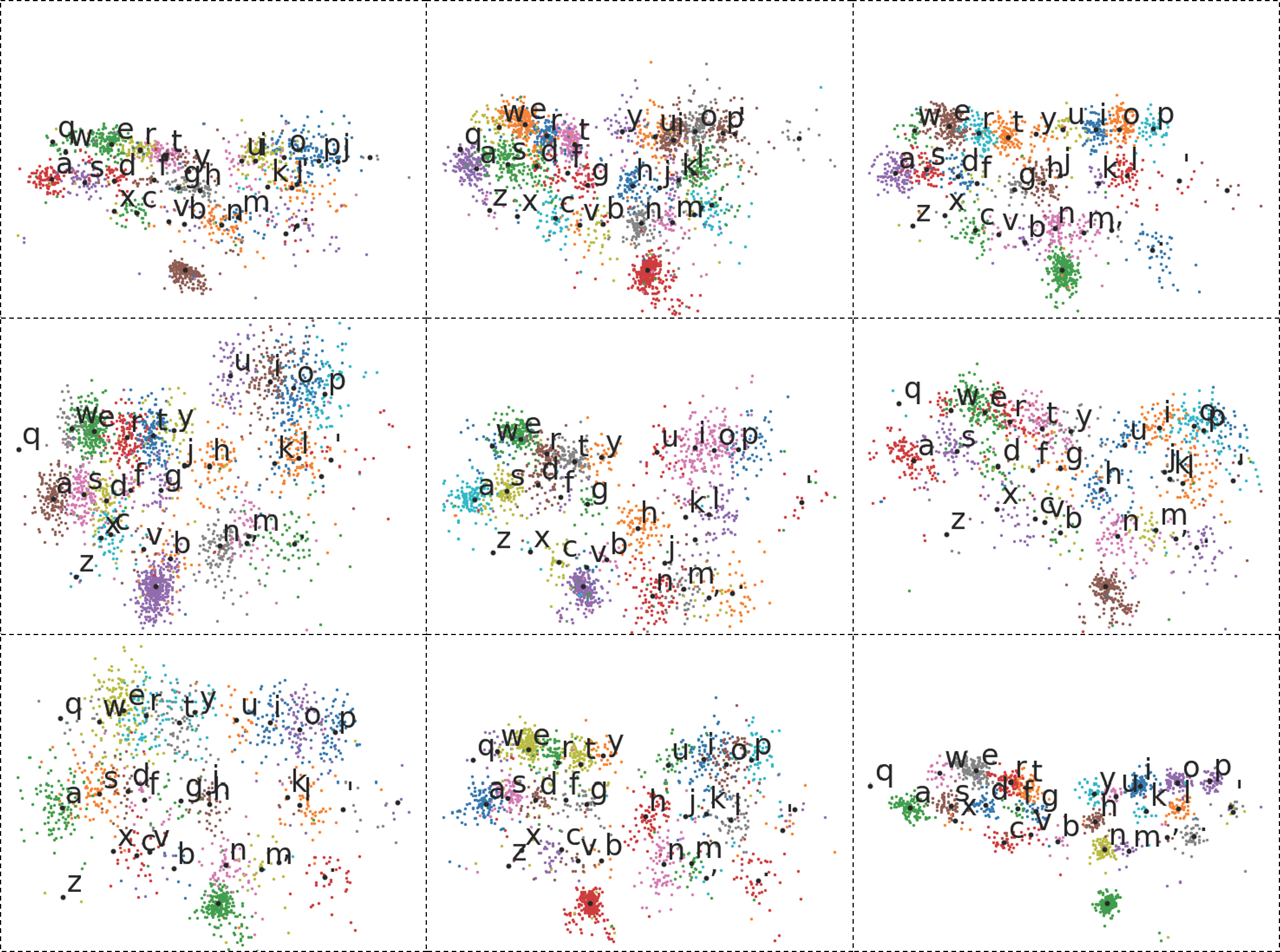}
	\caption{Examples of user mental models. The scales are normalized and the offsets are removed for visualization.}
	\label{fig:user_mental_model}
\end{figure}
Fig. \ref{fig:user_mental_model} exemplifies user mental models for the keyboard. We normalized the scales and removed the location offsets. Each user mental model characterizes different shape, size and location. Contrast to previous reports where users have typed on predefined regions, the keys do not align on straight lines, but rather on curves. We presume this results from the fact that the participants in this work have typed in a less-restricted environment. The participants in this study could naturally exhibit their typing behaviors with more freedom.

Though each user recognizes the keyboard layout in mentally different ways, each model consistently resembles the physical keyboard layout. This indicates that users can type on touch screens even though there is no visual and physical guidances.

\subsubsection{Physical Dimensions}
\begin{figure}
    \centering
    \subfloat[Variation of the scale in the horizontal direction]{
        \includegraphics[width=0.45\textwidth]{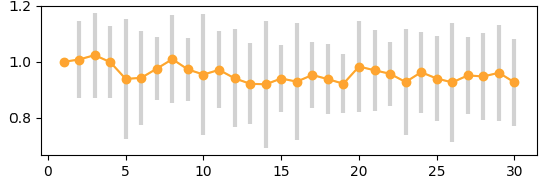}
        }\\
    \label{fig:user_behavior_scale_horizontal}
    ~ %add desired spacing between images, e. g. ~, \quad, \qquad, \hfill etc. 
      %(or a blank line to force the subfigure onto a new line)
    \subfloat[Variation of the scale in the vertical direction]{
        \includegraphics[width=0.45\textwidth]{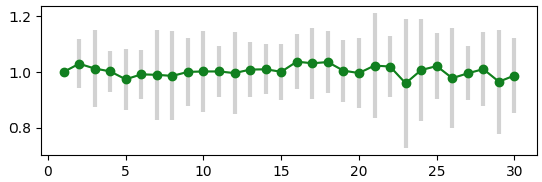}
        }
    \label{fig:user_behavior_scale_vertical}
    ~ %add desired spacing between images, e. g. ~, \quad, \qquad, \hfill etc. 
    %(or a blank line to force the subfigure onto a new line)
    \caption{The variation of scale over time. The scale does not change much in both horizontal and vertical directions indicating consistent user mental model size over time.}
    \label{fig:user_behavior_scale}
\end{figure}

\begin{table}
\centering
\caption{Summary of User Behavior Analysis}
\def\arraystretch{1.3}
\begin{tabular}{c | c | c | c | c}
\hline
\thickhline
Metric & Direction & Average & S.D. & Range \\
\thickhline
\multirow{2}{*}{{\bf{Scale}}} & Horizontal & $0.96$ & $0.16$ & $(0.92, 1.02)$ \\

& Vertical & $1.00$ & $0.13$ & $(0.96, 1.04)$ \\
\hline
\multirow{2}{*}{{\bf{Offset}}} & Horizontal & $44.73$ & $75.81$ & $[0.00, 70.07)$ \\

& Vertical & $25.95$ & $44.69$ & $[0.00, 45.57)$ \\
\hline
\multirow{2}{*}{{\bf{Size}}} & Horizontal & $259.0$ & $87.57$ & $(246.5, 276.2)$ \\

& Vertical & $125.9$ & $22.11$ & $(120.4, 130.4)$ \\
\hline
\thickhline
\end{tabular}
\label{tb:user_analysis_summary}
\end{table}
To estimate the physical dimensions, we utilized the distance between space and the character `p' in each sentence. We omitted the sentences that did not include `p' for the analysis. Table \ref{tb:user_analysis_summary} summarizes the analysis results. First, the average scale ratios among participants are 0.96 and 1.00 for horizontal and vertical directions, respectively. On average, users keep the overall dimension of the mental keyboard. The low standard deviations of the scales imply that the scales do not alter significantly over time. Fig. \ref{fig:user_behavior_scale} further examines the variation of the mental keyboard scales in both directions over time. As implied by the standard deviation, the scales do not vary greatly in both directions.

\begin{figure}
    \centering
    \subfloat[Variation of the offset in the horizontal direction]{
        \includegraphics[width=0.45\textwidth]{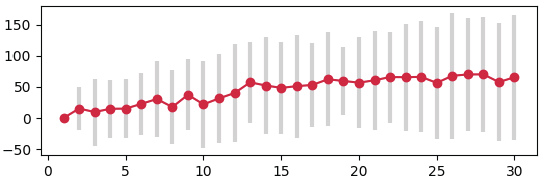}
        }\\
    \label{fig:user_behavior_offset_horizontal}
    ~ %add desired spacing between images, e. g. ~, \quad, \qquad, \hfill etc. 
      %(or a blank line to force the subfigure onto a new line)
    \subfloat[Variation of the offset in the vertical direction]{
        \includegraphics[width=0.45\textwidth]{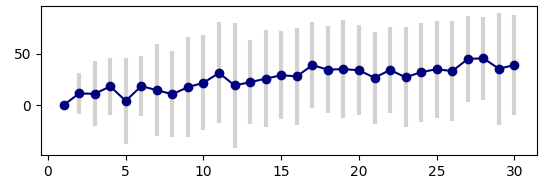}
        }
    \label{fig:user_behavior_offset_vertical}
    ~ %add desired spacing between images, e. g. ~, \quad, \qquad, \hfill etc. 
    %(or a blank line to force the subfigure onto a new line)
    \caption{The variation of offset over time. The offset increases in both horizontal and vertical directions indicating user mental model drifts over time.}
    \label{fig:user_behavior_offset}
\end{figure}

Next, the average offsets for horizontal and vertical directions are 44.73 and 25.95 pixels, respectively whereas the standard deviations are 75.81 and 44.69 for each direction, respectively. We calculated offsets compared to the mental keyboard model estimated from the first typed sentence. Users tend to move in both horizontal and vertical directions, though the tendency is stronger in the horizontal direction. Fig. \ref{fig:user_behavior_offset} investigates the change of the mental keyboard offsets over time. The mental models drift away from the initial position in both directions as expected. 

Lastly, the average size of the user mental models is smaller than the size of an actual physical keyboard ($350\times 150$ mm$^{2}$) in both directions. This contradicts the previous report which has stated that user mental models are slightly larger than an actual keyboard \cite{shi2018toast}. We surmise that the difference results from the different instructions provided to users and the different data collection settings. Compared to the previous work, we did not enforce any restrictions on the typing environment such as a pre-defined typing area or an initialization process for calibration. Thus, we expect our study can examine user behavior in a more natural and precise manner. We do not plot the variation of the size over time since it shows the same tendency as the scales with a different dimension (scale factor).

\subsubsection{Typing Feedback}
Table \ref{tb:typing_feedback} shows the user experience results depending on the feedback types. The results demonstrate that highlight feedback used in this study is more intuitive, convenient and satisfactory than the conventional asterisk feedback. When no feedback is given, users could hardly type the given text during the data collection process.

\begin{table}
\centering
\caption{Comparison of Three Types of Feedback}
\def\arraystretch{1.3}
\begin{tabular}{c | c | c | c}
\hline
\thickhline
Metric & None & Asterisk & Highlight \\
\thickhline
\bf{Intuitiveness} & $1.13$ & $3.63$ & $4.63$ \\

\hline
\bf{Convenience} & $1.5$ & $3.5$ & $4.25$ \\

\hline
\bf{Overall Satisfaction} & $1.25$ & $3.25$ & $4.13$ \\

\hline
\thickhline
\end{tabular}
\label{tb:typing_feedback}
\end{table}

\section{I-Keyboard System Design}
We describe the overall architecture of the proposed I-Keyboard and DND in this section. DND proposed in this paper allows the efficient and effective translation of user inputs into character sequences.

\subsection{System Architecture}
\begin{figure}
	\centering
	\includegraphics[scale=0.65]{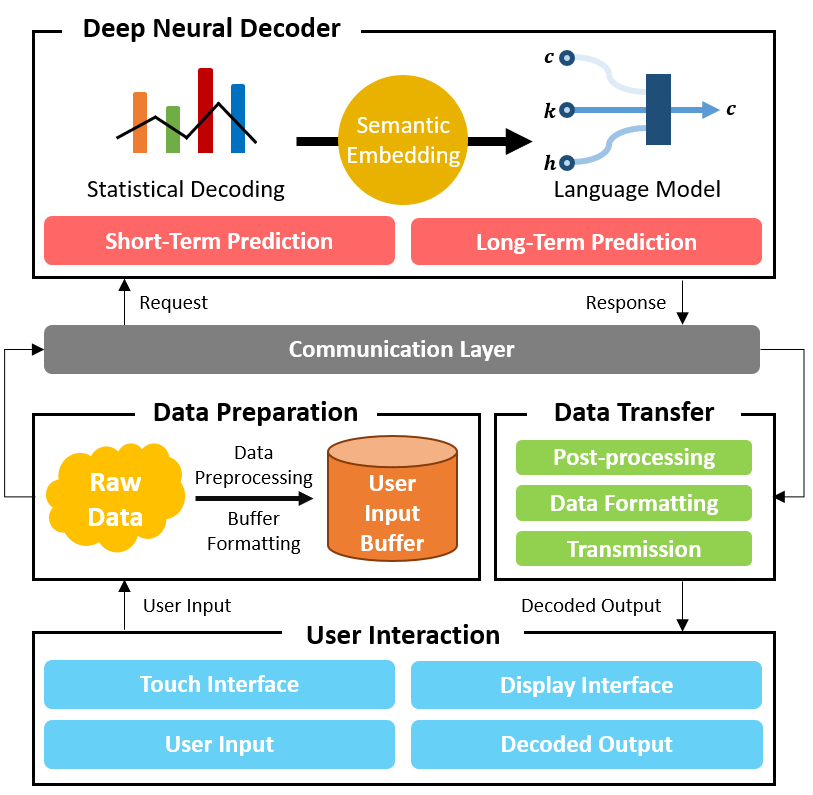}
	\caption{I-Keyboard system architecture. The user interaction module accepts user inputs and displays the decoding output. The data preparation and the data transfer modules function as a data pathway. DND translates user inputs into character sequences.}
	\label{fig:i-keyboard_architecture}
\end{figure}
Fig. \ref{fig:i-keyboard_architecture} describes the overall system architecture of I-Keyboard. The user interaction module receives user input through a touch interface and displays the decoded sequence through a display interface. The touch interface and the display interface can be contained in one device, though the present work separates the two. The data preparation module preprocesses and formats the raw user input and stores the data in a user input buffer. Since the proposed DND can consider both short-term and long-term dependencies present in the user data, the data from the past time steps improve the decoding performance. For this, the user input buffer keeps a certain number of past input data as a context. After the number of past input data gets stored in the user input buffer, the decoding becomes instant. The new input character is decoded with the past inputs stored in the user input buffer.

The communication layer between the data module and the DND module enables a tight integration of a deep learning framework and an application framework. Technically, current deep learning frameworks hardly support a native implementation of DL algorithms for application frameworks with a few exceptions. The need for the communication layer will disappear when the support for the native implementation emerges. DND decodes the user input in both short-term and long-term manners. When a user starts to type, DND decodes in the short-term manner due to the vacancy of user input buffer. After the user has been typing for a little while, DND uses the long-term prediction scheme. The details of DND follow in the next subsection. The data transfer module receives the decoded sequence and lets the interaction module display the result.

\subsection{Deep Neural Decoder (DND)}
\subsubsection{Problem Formulation}
We formulate the keyboard decoding problem as follows. Given a set of touch points on the touch screen $\mathbb{T} = (t_{1}, ..., t_{n})$ where $t_{i} = (x_{i}, y_{i})$ consisting of $x$ and $y$ positions and $n$ stands for the sequence length, a decoding algorithm finds the most probable sequence of characters $\hat{\mathbb{C}} = (\hat{c_{1}}, ..., \hat{c_{n}})$ where $\hat{c_i} \in D = \{e_{1}, ..., e_{k}\}$ for $1 \leq i \leq n$, $D$ is a character dictionary including space, enter, period and apostrophe, and $k$ is the size of the character dictionary:
\begin{equation}
    \hat{\mathbb{C}} = \mathop {\arg \max }\limits_{\mathbb{C}} {P(c_{1}, ..., c_{n}|t_{1}, ..., t_{n})} = \mathop {\arg \max }\limits_{\mathbb{C}} {P(\mathbb{C}|\mathbb{T})}.
    \label{eq:decoding_formulation}
\end{equation}
We use the maximum likelihood estimation (MLE). By applying the Bayes rule, the right hand side of (\ref{eq:decoding_formulation}) becomes
\begin{equation}
    \mathop {\arg \max }\limits_{\mathbb{C}} P(\mathbb{C}|\mathbb{T}) = \mathop {\arg \max }\limits_{\mathbb{C}} \frac{P(\mathbb{T}|\mathbb{C})\cdot P(\mathbb{C})}{P(\mathbb{T})}.
    \label{eq:decoding_formulation_bayes}
\end{equation}
Removing the invariant $P(\mathbb{T})$ turns the decoding objective into
\begin{equation}
    \hat{\mathbb{C}} = \mathop {\arg \max }\limits_{\mathbb{C}}{P(\mathbb{T}|\mathbb{C})\cdot P(\mathbb{C})}.
\end{equation}
In this setting, $P(\mathbb{T}|\mathbb{C})$ and $P(\mathbb{C})$ refer to a user mental model and language model, respectively. Since considering the long-term dependency complicates the statistical model for $P(\mathbb{T}|\mathbb{C})$, conventional models approximate the mental model by assuming the independence property and a Gaussian distribution. On the other hand, we model both the user mental model and the language model using bi-directional GRU's to handle the long-term dependency. By incorporating bi-directional GRU units, the proposed decoder can translate the input touch points into a character sequence without compromising the performance with the model complexity.

\subsubsection{Neural Architecture}
\begin{figure*}
	\centering
	\includegraphics[scale=0.9]{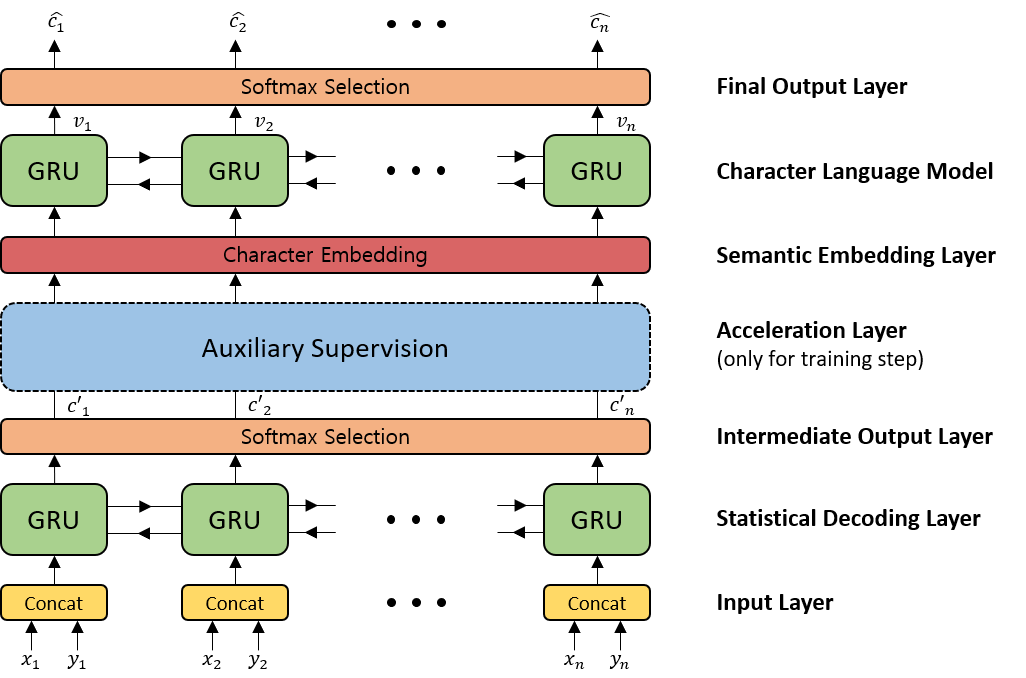}
	\caption{DND architecture. The statistical decoding layer extracts character sequences from positional information implicit to the user input. The character language model corrects positional errors with semantic information. The acceleration layer facilitates the learning process.}
	\label{fig:dnd_architecture}
\end{figure*}
Fig. \ref{fig:dnd_architecture} shows the neural architecture of DND. The user input flows from the input layer to the final output layer. The input layer concatenates the $x$ and $y$ components of user input and transforms the user input into input vectors. In addition, the input layer receives the user input at multiple time steps for the statistical decoding layer and the character language model (CLM) statistically and semantically examines both short-term and long-term dependencies. To support this, the input vectors are windowed and DND uses a certain number of past input vectors each time.

The statistical decoding layer receives the input vectors and translates them to a sequence of characters. The decoding layer does not take the semantic relations between characters into account, but the layer only considers positional relations among the user input during the translation. The decoding layer consists of a stack of bi-directional GRU units. Bi-directional GRU's can model the dependencies in forward and backward directions. Since I-Keyboard lets users type at any position on the touch interface, regarding both forward and backward dependencies is crucial for the decoding performance. We select GRU's instead of LSTM's for the sake of computational efficiency.

The softmax selection of the intermediate and the final output layers first maps the output from GRU units to a tensor whose dimension matches the size of the character dictionary. The softmax selection utilizes a linear layer for the mapping. Then, the softmax selection applies the softmax function to the tensor and selects the most probable components as follows:
\begin{equation}
    \hat{c_{i}}=\mathop {\arg \max }\limits_{c}\frac { exp(v_{i}^{c}) }{ \sum _{ j=1 }^{ k }{ exp(v_{i}^{j}) } },
\label{eq:softmax_selection}
\end{equation}
where $i$ ($1 \leq i \leq n$) is the index of the character, $v$ is the output vector of CLM and $k$ is the size of the character dictionary.

The acceleration layer enforces an additional auxiliary loss \cite{zhao2017pyramid} for the intermediate output to boost the training procedure. The acceleration layer significantly improves the convergence speed as well as the decoding accuracy. Without the auxiliary loss, the DND architecture becomes too deep for the gradient caused by the final loss to travel through the network. In such case, the learning would not occur in the appropriate direction for the middle layers. We use the softmax cross-entropy loss for both the acceleration layer and the final output loss. The acceleration layer is enabled only for the training step.

Next, the semantic embedding layer embeds each character into a character vector \cite{zhang2017understanding}. The embedding vectors stores the semantic relations between characters after training. The CLM revises the positional errors entailed in the user input data using the semantic relations between characters. It learns how characters are related and how to correct typos. The CLM also employs bi-directional GRU units to handle both forward and backward dependencies.

\section{Simulation}
In this section, we focus on verifying the decoding accuracy of the proposed DND compared to a couple of baselines. Then, we present ablation study results for various parameter choices.
\subsection{Simulation Setting}
\subsubsection{Baselines}

\begin{figure*}
	\centering
	\includegraphics[width=0.9\textwidth]{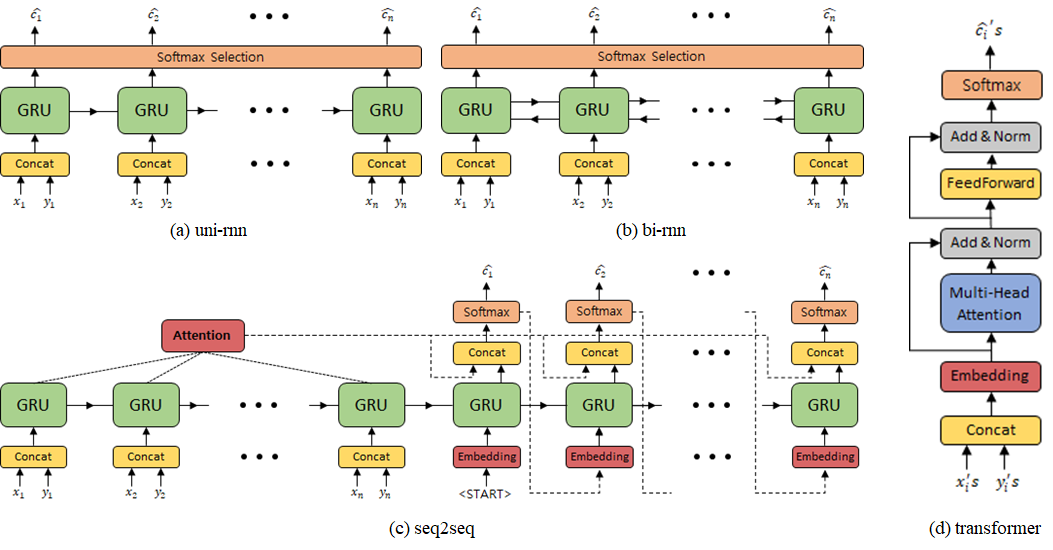}
	\caption{The architectures of the baselines. For the transformer model, we only employ the encoder part to generate a sequence of characters from user inputs. Other baselines do not vary from the basic architectures.}
	\label{fig:baseline_architectures}
\end{figure*}

We verified the performance of DND in comparison with a few baseline methods. We did not include conventional statistical models such as a Gaussian model \cite{zhu2018typing} since they failed to work in the challenging I-Keyboard scenario. The positions of user hands vary over time and the position of each key is not restricted to a specific area in the scenario. Therefore, each key can present at all positions of touch devices and the Gaussian modeling becomes meaningless. We selected four models among the latest and most popular deep learning networks \cite{8416973}: RNN (uni-rnn), bi-directional RNN (bi-rnn), encoder-decoder attention (seq2seq) \cite{luong2015effective} and transformer (trans) \cite{vaswani2017attention} models. Fig. \ref{fig:baseline_architectures} shows the architectures of the baseline models. For the transformer model, we only used the encoder part to transform the user inputs into the sequence of characters.

\iffalse
Fig. \ref{fig:baseline_architectures} depicts the simplified structure of each baseline model.
\fi

\subsubsection{Ablation Study}
We explored various parameter choices for each model. We modified multiple parameters and investigated the effect of parameter modifications on the decoding performance. We controlled the following parameters: the numbers of stacks (2, 3, 4), units for GRU (32, 64), heads (8, 16 for trans models) and the use of auxiliary loss (for DND).

\subsubsection{Data Preparation}
We separated the collected data into three parts: train, validation and test datasets. We randomly chose two participants and assigned their data as the test dataset. For the validation dataset, we selected one participant. We randomly shuffled the train dataset for better convergence. In addition, we augmented the train dataset by adding random offsets in both horizontal and vertical directions. We repeated the data augmentation process five times. We used the validation dataset to prevent overfitting. We stopped the training procedure when the error rate on the validation dataset started to increase.
\subsubsection{Evaluation Metrics}
We assessed each decoding algorithm with three metrics: decoding speed, character error rate (CER) and word error rate (WER). We measured the average decoding speed since ensuring a real-time operation is of great importance for decoders. We measured the accuracy with CER and WER. CER is defined as
\begin{equation}
    CER = \frac{MCD(S, P)}{length_{c}(P)} \times 100,
\end{equation}
where $MCD(S, P)$ is the minimum character distance between the decoded phrase $S$ and the ground-truth phrase $P$ and $length_{c}(P)$ is the number of characters in $P$. Similarly, WER is defined as
\begin{equation}
    WER = \frac{MWD(S, P)}{length_{w}(P)} \times 100,
\end{equation}
where $MWD(S, P)$ is the minimum word distance between $S$ and $P$ and $length_{w}(P)$ is the number of words in $P$. CER and WER count the number of insertions, deletions and substitutions of characters or words to transform $S$ into $P$.

\subsubsection{Implementation Details}
We implemented DND with Tensorflow (Python 3.6). We set the size of character embedding vector and the character dictionary as 16 and 31, respectively. We used Levenshtein measure to calculate MCD and MWD. We set other model parameters as stated in Section V-A-2). For training, we used Adam optimizer and set the initial learning rate as 0.001. We increased the learning rate when the minimum validation loss was updated and decreased the learning rate when the minimum validation loss was not updated. The rate of learning rate variation was 0.1. We stopped training when the minimum validation loss did not change for 10 epochs. We chose models with the minimum validation loss for the simulation. We used one GPU (Titan X) for training and one CPU (2.9 GHz Intel Core i7) on MacBook Pro for testing.

\subsection{Results and Analysis}
\begin{table}
\centering
\caption{Summary of Simulation Results}
\def\arraystretch{1.3}
\begin{tabular}{c | c | c | c | c}
\hline
\thickhline
Model Name & Parameter & CER & WER & Time \\
\thickhline
\multirow{6}{*}{{\bf{uni-rnn}}} & s2u32 & 11.21 & 41.27 & 0.72 msec \\
& s2u64 & 7.93 & 30.89 & 0.95 msec \\
& s3u32 & 9.19 & 35.32 & 0.94 msec \\
& s3u64 & 7.10 & 28.53 & 0.89 msec \\
& s4u32 & 7.15 & 29.22 & 1.12 msec \\
& s4u64 & 6.42 & 25.21 & 1.08 msec \\
\hline
\multirow{6}{*}{{\bf{bi-rnn}}} & s2u32 & 6.22 & 22.85 & 0.88 msec \\
& s2u64 & 4.53 & 16.62 & 0.98 msec \\
& s3u32 & 4.08 & 15.24 & 1.11 msec \\
& s3u64 & 1.84 & 8.73 & 1.10 msec \\
& s4u32 & 4.08 & 16.62 & 1.28 msec \\
& s4u64 & 1.99 & 9.97 & 1.33 msec \\
\hline
\multirow{6}{*}{{\bf{seq2seq}}} & s2u32 & 83.45 & 100 & 1.97 msec \\
& s2u64 & 54.43 & 68.98 & 2.04 msec \\
& s3u32 & 64.18 & 81.44 & 2.54 msec \\
& s3u64 & 15.82 & 18.14 & 2.58 msec \\
& s4u32 & 48.92 & 63.16 & 2.63 msec \\
& s4u64 & 8.59 & 9.97 & 2.69 msec \\
\hline
\multirow{4}{*}{{\bf{trans}}} & s3u32h8 & 24.91 & 56.37 & 0.65 msec \\
& s3u32h16 & 28.16 & 64.68 & 0.71 msec \\
& s3u64h8 & 23.70 & 55.26 & \bf{0.63 msec} \\
& s3u64h16 & 23.98 & 55.40 & 0.74 msec \\
\iffalse
& s4u32h8 & cer & wer & time \\
& s4u32h16 & cer & wer & time \\
& s4u64h8 & cer & wer & time \\
& s4u64h16 & cer & wer & time \\
\fi
\hline
\multirow{12}{*}{{\bf{DND}}} & s2u32 & 77.33 & 100 & 1.67 msec \\
& s2u32au & 2.37 & 9.00 & 1.67 msec \\
& s2u64 & 68.19 & 100 & 1.72 msec \\
& s2u64au & \bf{1.18} & \bf{4.16} & 1.65 msec \\
& s3u32 & 76.22 & 100 & 2.05 msec \\
& s3u32au & 1.86 & 6.65 & 2.12 msec \\
& s3u64 & 75.26 & 100 & 2.06 msec \\
& s3u64au & 1.54 & 5.12 & 2.10 msec \\
& s4u32 & 75.77 & 100 & 2.72 msec \\
& s4u32au &4.33 & 15.51 & 2.50 msec \\
& s4u64 & 77.36 & 100 & 2.65 msec \\
& s4u64au & 1.26 & 4.43 & 2.72 msec \\
\hline
\thickhline
\end{tabular}
\label{tb:simulation_summary}
\end{table}

Table \ref{tb:simulation_summary} summarizes the simulation results. The characters `s', `u', `h', and `au' in the parameter column indicate the numbers of stacks, units of GRU and heads and the presence of auxiliary loss, respectively and each number following the character specifies the dimension. CER, WER and time (the average time needed to decode one word) in the table were evaluated on the test data. Table \ref{tb:simulation_examples} displays decoding examples of bi-rnn and DND models compared to the ground-truth. 

Comparing the performance between models reveals the intuitions for designing DND. First of all, the results of uni-rnn and bi-rnn imply the importance of considering the backward dependencies in decoding performance. The superior performance of bi-rnn (s2u32, s3u32, s4u32) over uni-rnn (s2u64, s3u64, s4u64) proves the importance of the backward dependency since uni-rnn with 64 GRU units and bi-rnn with 32 GRU units have approximately the same model size. In addition, uni-rnn could not achieve high accuracy even when the model size increases.

Next, the superior performance of DND over bi-rnn suggests the limitation of decoding with only positional information. Though bi-rnn models showed reasonable performance, they could not correct the positional mistakes made by users. By employing semantic embedding, DND further improved the performance of bi-rnn. In addition, the auxiliary loss forced by the acceleration layer lets gradients pass through the DND architecture. Without the auxiliary loss, DND becomes too deep for the single loss to train. The performance difference between DND with the auxiliary loss and DND without the loss verifies the effectiveness of the acceleration layer.

The seq2seq and trans models did not exhibit acceptable performance although they have demonstrated the state of the art performance in other tasks such as neural translation. We presume the seq2seq and trans models require semantic information implicit in the input data to show high performance as reported. The input for the decoding task in this work rarely contained semantic information but only physical (positional) information. The lack of semantic meaning in the input data hindered the models to translate the input sequence to the character sequence.

All of the models guarantee a real-time operation with reasonable decoding speeds. We did not use a special computation device, but just a normal one. Moreover, we did not utilize a batch-based computation since DND accepts user input one at a time, not in the batch-based form in the real-world setting. When DND accepts a batch of input, the throughput increases several times (up to the batch size). If DND is deployed in a server, the server can utilize the batch-based computation.

\begin{table}
\centering
\caption{Examples of Decoding Results}
\def\arraystretch{1.3}
\begin{tabular}{c | c | l}
\hline
\thickhline
\multirow{3}{*}{{\bf{Ex. 1}}} & G.T. & do you know where there's a store that sells towels.\\
& bi-rnn & do you know where there's a store that sells towels.\\
& DND & do you know where there's a store that sells towels.\\

\hline
\multirow{3}{*}{{\bf{Ex. 2}}} & G.T. & can i have a receipt please.\\
& bi-rnn & can i have a receilt please.\\
& DND & can i have a receipt please.\\

\hline
\multirow{3}{*}{{\bf{Ex. 3}}} & G.T. & my throat is sore.\\
& bi-rnn & my thrat his sore.\\
& DND & my theat ois sore.\\

\hline
\thickhline
\end{tabular}
\label{tb:simulation_examples}
\end{table}

\section{Experimental Verification}
To demonstrate the performance of I-Keyboard in real application settings, we design an experiment incorporating users. We report the typing speed and the accuracy as performance metrics.
\subsection{Experiment Setting}
\subsubsection{Participants}
13 skilled QWERTY keyboard users (5 male, 8 female) from the campus who had not joined in the data collection process participated in the experiments. Their ages range from 22 to 32 (average = 25.6, std = 2.5). All subjects responded that they used on daily basis both physical keyboards and soft keyboards on mobile devices and they could type in an eyes-free manner.
\subsubsection{Apparatus}
We conducted the experiments using the same apparatus of the data collection process. In addition, we measured the time spent on typing the task phrases. Other than that, we used the same interface as in the data collection procedure.
\subsubsection{Baselines}
We did not select a baseline method for comparison in the experiments for the following reasons. First, we pioneered the I-Keyboard scenario in which users type using ten fingers in an eyes-free manner with no pre-defined region for typing and no calibration step. As this scenario is first attempted in this work, this type of method has not been mentioned in the literature. Second, there is no probability that conventional methods with an enhanced calibration may perform better. The I-Keyboard setting reduces the constraints other methods impose such as a pre-defined typing region and a calibration step. Without such constraints, the decoding becomes statistically more difficult since the touch points spread over larger area and vary significantly point after point due to hand position shifts. Third, we designed I-Keyboard through comprehensive simulations which already compared various model architectures and parameters. We only referred to the performance metrics of the previous state of the art \cite{shi2018toast} to assess the performance of I-Keyboard, since the accuracy of the state of the art in the challenging I-Keyboard setting was only 32.37\% (CER).
\subsubsection{Performance Metrics}
We evaluated the I-Keyboard performance with the following metrics: speed in WPM, accuracies in CER and WER, and subjective rates in 5-Likert scale. WPM is defined as follows:
\begin{equation}
    WPM = \frac{|length_{c}(S) - 1|}{M} \times \frac{1}{n_{c}},
\end{equation}
where $M$ is the time elapsed from the first key stroke to the last key stroke for $S$ in minutes and $n_{c}$ is the average number of characters in a word. We set $n_{c} = 5$ in the analysis. We gathered subjective feedback from the participants in five perspectives: mental, physical and temporal demands, comfortability, and overall satisfaction.
\subsubsection{Procedure}
We first collected the demographic information from the participants. Then, we informed them the experiment procedure. The experiments consisted of two phases: typing with a physical keyboard and I-Keyboard. To remove the effect of bias, we alternated the order of typing devices from participant to participant. We instructed the participants to type as accurately and quickly as possible. For each keyboard, we first presented 10 phrases randomly sampled from the phrase set as a warm up session. After participants became familiar with each keyboard, we required each participant to transcribe 20 phrases randomly selected from the phrase set. Finally, participants took a survey to rate I-Keyboard subjectively.

\subsection{Results and Analysis}
\begin{figure}
	\centering
	\includegraphics[scale=1]{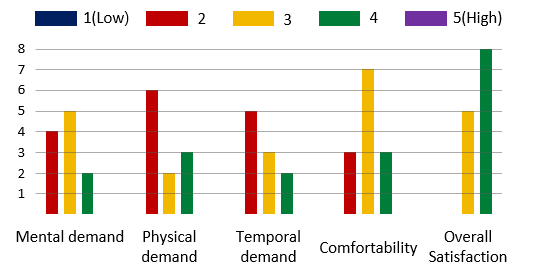}
	\caption{Subjective ratings from the participants. The result attests that users can type with little mental and physical fatigue.}
	\label{fig:subjective_ratings}
\end{figure}

Fig. \ref{fig:subjective_ratings} and Table \ref{tb:comments_from_users} summarize the experiment result. The average typing speeds of the physical keyboard and I-Keyboard were 45.57 and 51.35 WPM, respectively. The average speed of the physical keyboard measured in this work is slower than the previously reported average speed (55 WPM) \cite{shi2018toast}. We investigated the typing speed of each participant and some of them showed fairly slow typing speeds with both the physical keyboard and I-Keyboard. The minimum typing speeds for each typing interface are 35.13 and 26.88 WPM, respectively. This factor lowers the average typing speeds of both interfaces. However, the typing speed of I-Keyboard is faster than the previous state of the art (41.4 WPM), though the average speed of I-Keyboard might have been reduced.

In addition, we report much higher speed ratio between the typing interfaces (45.57/51.35 = 88.74\%) than the baseline (41.4/55.5 = 74.59\%). We infer from user feedback that the enhanced typing speed resulted from the increased freedom provided by I-Keyboard. Users can type anywhere on the touch screen without worrying about key positions and typos. Participants replied that this feature makes I-Keyboard convenient. In fact, we can further improve the typing speed of I-Keyboard with more sensitive touch devices. Participants tried to type faster but the device in this work could not sense the touches when the typing speed exceeded the sensitivity of the device. We expect I-Keyboard would offer more comfortable and faster typing interface with careful choice of touch device.

The overall accuracies of I-Keyboard in CER and WER are 98.91\% and 96.12\%, respectively. The accuracy is slightly higher than the accuracy of the test set in simulation. We assume the participants attempted to type as precisely as possible following the instruction given before the experiment. We instructed the participants to type as precisely and swiftly as possible. For the data collection process, we did not inform such instruction.

The subjective rating from the participants reveals key features of I-Keyboard. The physical and temporal demands are left-skewed, meaning users can type quickly without experiencing physical fatigue. This implies improved typing speed and user experience. The mental demand and comfortability approximately show Gaussian distributions signifying these metrics require further investigation to extract meaning. We suspect the participants do not take typing as a burden. The overall satisfaction is 3.62.

\begin{table*}
\centering
\caption{Comments from users}
\def\arraystretch{1.3}
\begin{tabular}{c | l}
\hline
\thickhline
\multirow{3}{*}{\bf{Positive}}
& ``I expect I-Keyboard will show more excellent performance in terms of speed and comfort with better touch device." \\
& ``It was nice that I can comfortably and freely position my hands." \\
& ``I could hit the keyboard with my usual typing habit instead of typing it on the pre-defined keyboard."\\
\hline
\multirow{3}{*}{\bf{Negative}} 
& ``It would be better if the touch device can perceive fast touches" \\
& ``Users can get confused with the keys in similar locations." \\
& ``I-Keyboard could not recognize key press when I touch it with my fingernail."\\
\hline
\thickhline
\end{tabular}
\label{tb:comments_from_users}
\end{table*}

\section{Discussion}
I-Keyboard implements an eyes-free ten finger typing interface that does not involve a calibration procedure. Moreover, users do not need to learn any new concept regarding I-Keyboard prior to usage. They can just start typing naturally by transferring the knowledge of physical keyboards. Users can keep typing even when they have taken off their hands phrase after phrase without an additional calibration step. I-Keyboard targets mobile devices since most mobile devices (Android iOS) support QWERTY keyboards and users in general use them. However, there still exist a few remaining works for further improvements although we established the effectiveness of I-Keyboard.

First of all, we can further experiment with various touch screen sizes. In this paper, we utilized a relatively large touch screen. By using the large touch screen, we intended to consider radical situations such as moving hands dynamically phrase after phrase. Even in such situations, I-Keyboard can decode the key strokes with high accuracy. Thus, users can both naturally and comfortably type with I-Keyboard. Furthermore, increasing display size is the trend for mobile devices and the trend will accelerate with the development of foldable screens \cite{wu2018high}. The current version of I-Keyboard can already support smartphones with a few adjustments such as supporting typing on a QWERTY keyboard with two thumbs. Nevertheless, we plan to extend I-Keyboard to other touch screens in the succeeding study. When targeting multiple screen sizes, transfer learning \cite{wang2019softly} would enable one general platform for all.

We implemented I-Keyboard with the LG touch monitor using WPF. In the following research, we can extend the presented I-Keyboard environment to any flat surfaces with advanced touch sensing techniques. A depth-based touch detection system \cite{wilson2010using} could enable such extension. When I-Keyboard is integrated with advanced touch detection methods, a number of mobile systems would benefit from it. For example, I-Keyboard can offer an effective text-entry method for VR systems. Since VR systems require an eyes-free text-entry method, I-Keyboard can be an attractive alternative for current text-entry tools for VR systems.

Finally, the next generation I-Keyboard can support non-alphabetic characters (e.g., numbers, punctuation and functional keys) for better user experience, though users can type lower case alphabets, apostrophe, enter, space and period with the current I-Keyboard. Specifically, the shift key can be detected by a multi-touch detection algorithm by recognizing the first key pressed while the second key is stroke as the shift key. A gesture detection algorithm would enable the delete key by assigning a swipe gesture as the delete key. In addition, users would not need to wear gloves with a multi-touch detection algorithm by detecting palms. We plan to develop the above-mentioned features for the next generation I-Keyboard.

\section{Conclusion}
In this paper, we proposed I-Keyboard with DND. I-Keyboard, for the first time, attempted to realize a truly imaginary keyboard that does not require calibration. Users can start typing from anywhere on the touch screen without concerning about the keyboard shape and location in an eyes-free manner. In addition, users do not need to learn anything before typing. For the development of I-Keyboard, we conducted a user study while collecting the largest dataset. We analyzed the user behaviors in the eyes-free ten-finger scenario which enforces minimum constraints. User mental models consistently displayed similar layouts as a physical keyboard, though the models drift in location and vary in shape over time. After confirming the feasibility, we designed I-Keyboard and DND. We utilized a deep neural architecture to handle the dynamic variation of user mental models and the semantic embedding technique to further boost the decoding performance. Simulations and experiments verified the superior performance of I-Keyboard. The accuracy of DND was 95.84\% and 96.12\% in the simulation and the experiment, respectively surpassing the performance of the baseline by 4.06\%. Users can type at 45.57 WPM (88.74\% of the typing speed with a physical keyboard) breaking the previous state of the art performance (41.4 WPM, 74.6\%). 

% if have a single appendix:
%\appendix[Proof of the Zonklar Equations]
% or
%\appendix  % for no appendix heading
% do not use \section anymore after \appendix, only \section*
% is possibly needed

% use appendices with more than one appendix
% then use \section to start each appendix
% you must declare a \section before using any
% \subsection or using \label (\appendices by itself
% starts a section numbered zero.)
%

%
%\appendices
%\section{Proof of the First Zonklar Equation}
%Appendix one text goes here.
%
%\section{}
%Appendix two text goes here.
%

% use section* for acknowledgment
%\section*{Acknowledgment}
%The authors would like to thank...

% Can use something like this to put references on a page
% by themselves when using endfloat and the captionsoff option.
\ifCLASSOPTIONcaptionsoff
  \newpage
\fi

\begin{IEEEbiography}[{\includegraphics[width=1in,height=1.25in,clip,keepaspectratio]{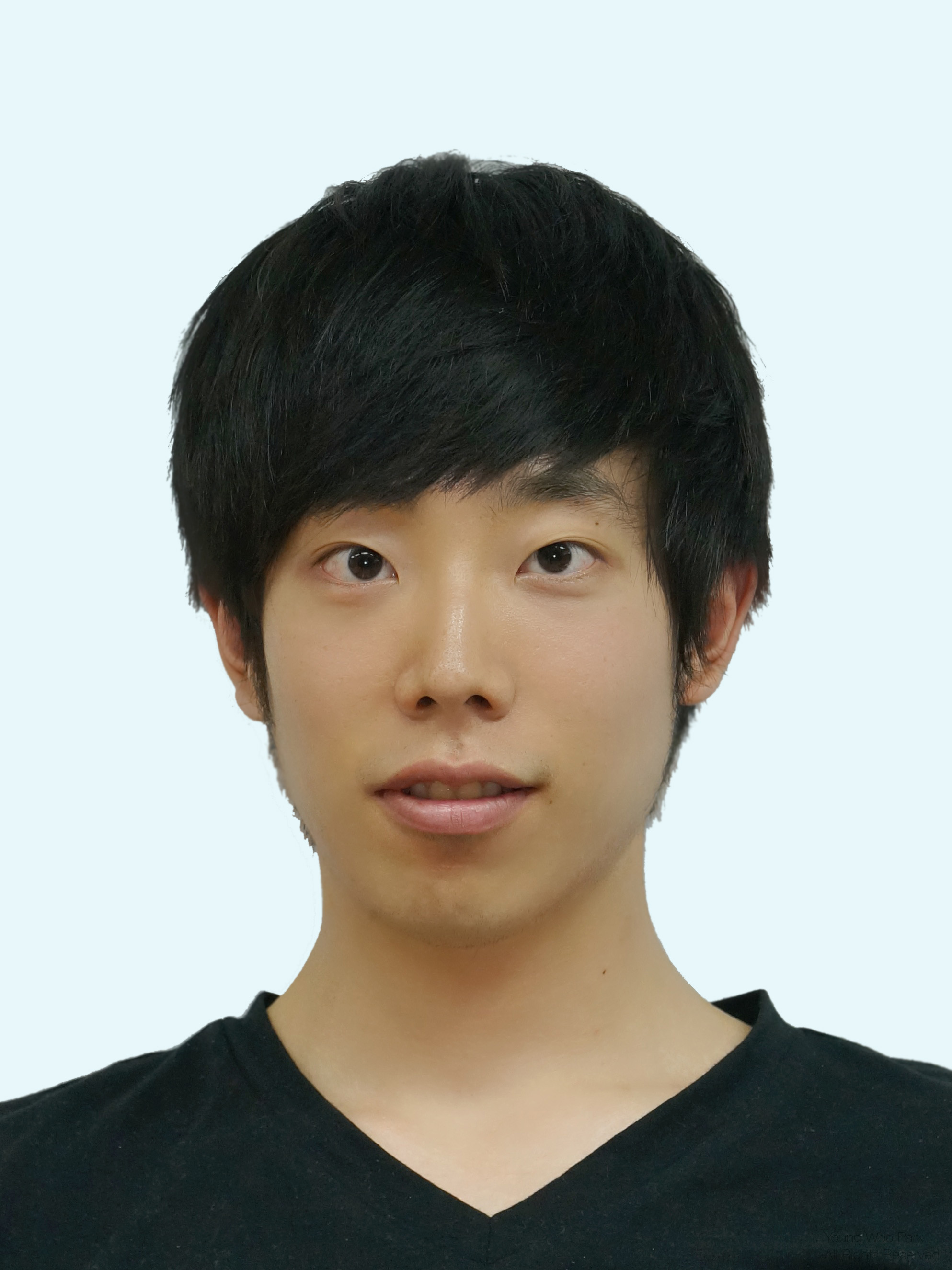}}]{Ue-Hwan Kim}
% or if you just want to reserve a space for a photo:
received the M.S. and B.S. degrees in Electrical Engineering from Korea Advanced Institute of Science and Technology (KAIST), Daejeon, Korea, in 2015 and 2013, respectively. He is currently pursuing the Ph.D. degree at KAIST. His current research interests include service robot, human robot interaction, cognitive IoT, human computer interaction, computational memory systems and learning algorithms.
\end{IEEEbiography}

\begin{IEEEbiography}[{\includegraphics[width=1in,height=1.25in,clip,keepaspectratio]{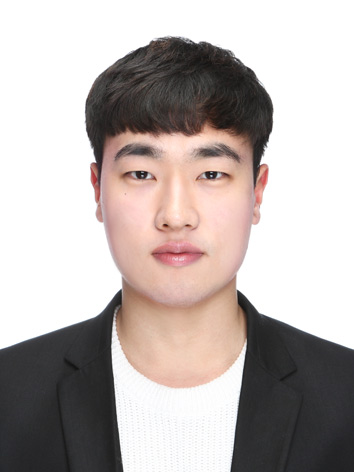}}]{Sahng-Min Yoo}
received his B.S degree in Electrical Engineering from Korea Advanced Institute of Science and Technology (KASIT), Daejeon, Korea, in 2017. He is currently working toward integrated Master's and Doctoral Degree at KAIST. His current research interests include user experience, service robot and deep learning algorithms.
\end{IEEEbiography}

\begin{IEEEbiography}[{\includegraphics[width=1in,height=1.25in,clip,keepaspectratio]{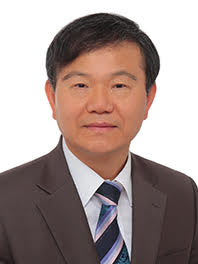}}]{Jong-Hwan Kim}
% or if you just want to reserve a space for a photo:
(F'09) received the Ph.D. degree in electronics engineering from Seoul National University, Korea, in 1987. Since 1988, he has been with the School of Electrical Engineering, KAIST, Korea, where he is leading the Robot Intelligence Technology Laboratory as KT Endowed Chair Professor. Dr. Kim is the Director for both of KoYoung-KAIST AI Joint Research Center and Machine Intelligence and Robotics Multi-Sponsored Research and Education Platform. His research interests include intelligence technology, machine intelligence learning, and AI robots. He has authored 5 books and 5 edited books, 2 journal special issues and around 400 refereed papers in technical journals and conference proceedings.
\end{IEEEbiography}

% You can push biographies down or up by placing
% a \vfill before or after them. The appropriate
% use of \vfill depends on what kind of text is
% on the last page and whether or not the columns
% are being equalized.

%\vfill

% Can be used to pull up biographies so that the bottom of the last one
% is flush with the other column.
%\enlargethispage{-5in}

\end{document}